\documentclass[aps,pre,twocolumn,groupedaddress]{revtex4-1}
\usepackage{graphicx}
\usepackage{amsmath}
\usepackage{amsfonts}
\usepackage{color}
\definecolor{Blue}{rgb}{0,0,0.6}
\definecolor{Red}{rgb}{0.6,0,0}

\begin{document}

%\preprint{}

\title{Leaf-to-leaf distances and their moments in finite and infinite ordered $m$-ary tree graphs}


\author{Andrew M.\ Goldsborough}
\email[]{a.goldsborough@warwick.ac.uk}
\homepage[]{\\www.warwick.ac.uk/andrewgoldsborough}
%\thanks{}
\affiliation{Department of Physics and Centre for Scientific Computing, The University of Warwick, Coventry, CV4 7AL, United Kingdom}

\author{S.\ Alex Rautu}
\email[]{s.a.rautu@warwick.ac.uk}
%\homepage[]{http://www2.warwick.ac.uk/fac/sci/physics/current/postgraduate/pglist/phrhby}
%\thanks{}
\affiliation{Department of Physics and Centre for Scientific Computing, The University of Warwick, Coventry, CV4 7AL, United Kingdom}

\author{Rudolf A.\ R\"{o}mer}
\email[]{r.roemer@warwick.ac.uk}
\homepage[]{www.warwick.ac.uk/rudoroemer}
%\thanks{}
\affiliation{Department of Physics and Centre for Scientific Computing, The University of Warwick, Coventry, CV4 7AL, United Kingdom}


\date{\today}

\begin{abstract}
We study the leaf-to-leaf distances on one-dimensionally ordered, full and complete $m$-ary tree graphs using a recursive approach. In our formulation, unlike in traditional graph theory approaches, leaves are ordered along a line emulating a one dimensional lattice. We find explicit analytical formulae for the sum of all paths for arbitrary leaf separation $r$ as well as the average distances and the moments thereof. We show that the resulting explicit expressions can be recast in terms of Hurwitz-Lerch transcendants. Results for periodic trees are also given. For incomplete random binary trees, we provide first results by numerical techniques; we find a rapid drop of leaf-to-leaf distances for large $r$.
\end{abstract}

%\keywords{}

\maketitle

\section{Introduction}
%%%%%%%%%%%%%%%%%%%%%%%%%%%%%%%%%%%%%%%%%%%%%%%%%%%%%%%%%%%%%%%%%%%
The study of graphs and trees, i.e.\ objects (or \textit{vertices}) with pairwise relations (or \textit{edges}) between them, has a long and distinguished history throughout nearly all the sciences. In computer science, graphs, trees and their study are closely connected, e.g.\ with sorting and search algorithms \cite{SedF13}; in chemistry the Wiener number is a topological index intimately correlated with, e.g., chemical and physical properties of alkane molecules \cite{Wie47}. In physics, graphs are equally ubiquitous, not least because of their immediate usefulness for systematic perturbation calculations in quantum field theories \cite{PesS95}. In mathematics, graph theory is in itself an accepted branch of mainstream research and graphs are a central part of the field of discrete mathematics \cite{Ros12}.
An important concept that appears in all these fields is the \textit{distance} in a graph, i.e.\ the number of edges connecting two vertices \cite{LecNS13,SzeWW11,Wan10}. For trees, i.e.\ undirected graphs in which any two vertices are connected by only one path,
various results exist \cite{Jac88,KirPS89,KirPS94}, for example, that compute the distance from the top of the tree to its leaves.
%In a binary tree such as shown in Fig.\ \ref{fig-binarytree} this distance might correspond, e.g.\ to the number of yes/no decisions one performs when searching for information.

Tree-like structures have recently also become more prominent in quantum physics of interacting particles with the advent of so-called tensor network methods \cite{Sch11}. These provide elegant and powerful tools for the simulation of low dimensional quantum many-body systems. In a recent publication \cite{GolR14} we show that certain correlation functions and measures of quantum entanglement can be constructed by a holographic distance and connectivity dependence along a tree network connecting certain leaves \cite{EveV11}. In these quantum systems, the leaves are ordered according to their physical position, for example the location of magnetic ions in a quantum wire. This ordering imposes a new restriction on the tree itself and the lengths which become important are leaf-to-leaf distances across the ordered tree. We emphasize that these distances therefore correspond to quite different measures than those studied in the various sciences mentioned before. We also note that in tensor networks the leaf-to-leaf distance is referred to as the \textit{path length} \cite{EveV11}, but in graph theory this term usually refers to the sum of the levels of each of the vertices in the tree \cite{SedF13}.

\begin{figure*}[t]
(a) \includegraphics[width=0.95\columnwidth]{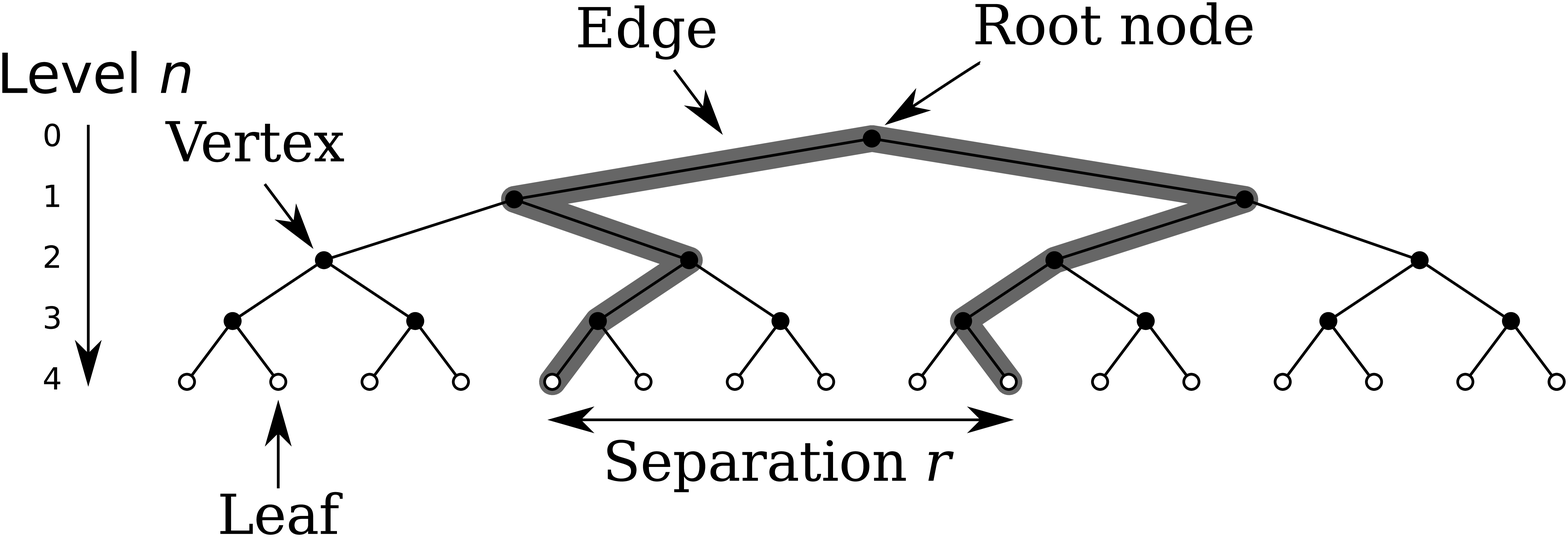}
(b) \includegraphics[width=0.95\columnwidth]{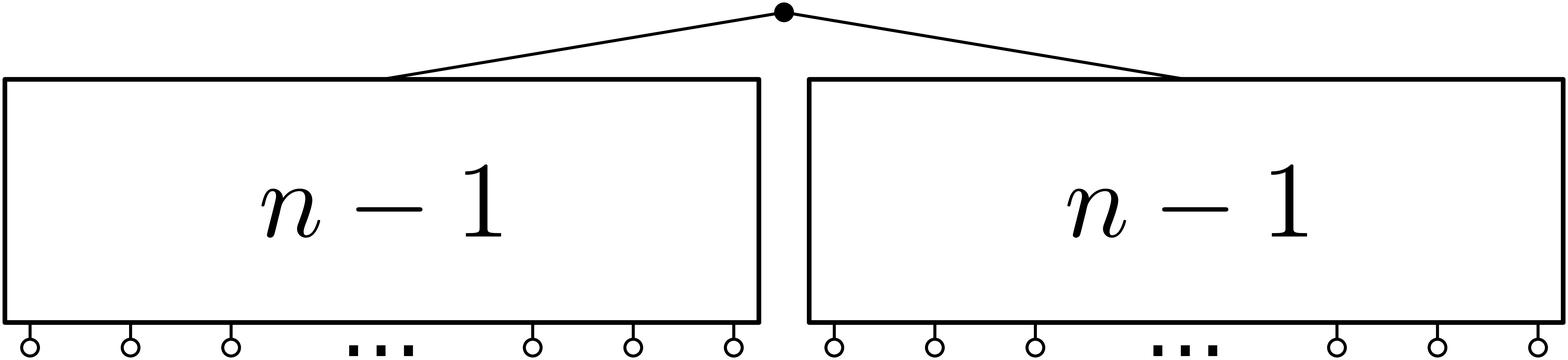}
\caption{(a) A complete binary tree with various definitions discussed in main text labeled. Circles ($\bullet$, $\circ$) denote vertices while lines indicate edges between the vertices of different depth. The tree as shown has a depth of $4$ and $L=16$ leaves ($\circ$). The indicated separation is $r=5$ while the associated leaf-to-leaf distance equals $\ell=8$ as indicated by the thick line.%
(b) Schematic decomposition of a level $n$ tree with root node ($\bullet$) and leaves ($\circ$) into two level $n-1$ trees (rectangles) each of which has $2^{n-1}$ leaves.}
\label{fig-binarytree-maximalpaths}
\label{fig-binarytree}
\end{figure*}

In the present work, we shall concentrate on full and complete trees that have the same structure as regular tree tensor networks \cite{SilGMR10,GerSRF14}. We derive the average leaf-to-leaf distances for varying leaf separation with leaves ordered in a one-dimensional line as shown e.g.\ in Fig.\ \ref{fig-binarytree}(a) for a binary tree \footnote{This is the information needed by the holography methods used in Ref.\ \cite{GolR14}.}.
%%%%%%%%%%%%%%%%%%%%%%%%%%%%%%%%%%%%%%%%%%%%%%%%%%%%%%%%%%%%%%%%%%%
%%%%%%%%%%%%%%%%%%%%%%%%%%%%%%%%%%%%%%%%%%%%%%%%%%%%%%%%%%%%%%%%%%%
The method is then generalized to $m$-ary trees and the moments of the leaf-to-leaf distances. Explicit analytical results are derived for finite and infinite trees. We also consider the case of periodic trees. We then illustrate how such properties may arise in the field of tensor networks. Last, we numerically study the case of incomplete random trees, which is closest related to the tree tensor networks considered in Ref.\ \cite{GolR14}.

%%%%%%%%%%%%%%%%%%%%%%%%%%%%%%%%%%%%%%%%%%%%%%%%%%%%%%%%%%%%%%%%%%%
\section{Average leaf-to-leaf distance in complete binary trees}
\label{sec-binary-averagepathlength}
%%%%%%%%%%%%%%%%%%%%%%%%%%%%%%%%%%%%%%%%%%%%%%%%%%%%%%%%%%%%%%%%%%%

%%%%%%%%%%%%%%%%%%%%%%%%%%%%%%%%%%%%%%%%%%%%%%%%%%%%%%%%%%%%%%%%%%%
\subsection{Recursive formulation}
\label{sec-binary-averagepathlength-recursion}
%%%%%%%%%%%%%%%%%%%%%%%%%%%%%%%%%%%%%%%%%%%%%%%%%%%%%%%%%%%%%%%%%%%

Let us start by considering the complete binary tree shown in Figure \ref{fig-binarytree}(a). It is a connected graph where each vertex is $3$-valent and there are no loops. The \textit{root node} is the vertex with just two degrees at the top of Figure \ref{fig-binarytree}(a).
The rest of the vertices each have two \textit{child} nodes and one parent. A \textit{leaf node} has no children.
The \textit{depth} of the tree denotes the number of vertices from the root node with the root node at depth zero. With these definitions, a binary tree is \textit{complete} or \textit{perfect} if all of the leaf nodes are at the same depth and all the levels are completely filled.
We now denote by the \textit{level}, $n$, a complete set of vertices that have the same depth. These are enumerated with the root level as $0$. We will refer to a \textit{level $n$ tree} as a complete tree where the leaves are at level $n$. The \textit{leaf-to-leaf distance}, $\ell$, is the number of edges that are passed to go from one leaf node to another (cp.\ Figure \ref{fig-binarytree}(a)).

Let us now impose an \textit{order} on the tree of Figure \ref{fig-binarytree}(a) such that the leaves are enumerated from left to right to indicate position values, $x_i$, for leaf $i$. Then we can define a leaf \textit{separation} $r=|x_{i} - x_{j}|$ for any pair of leaves $i$ and $j$. This is equivalent to the notion of distance on a one-dimensional physical lattice.
Let the \textit{length} $L$ be the length of the lattice, i.e.\ number of leaf nodes. Then for such a complete binary tree, we have $L = 2^{n}$.

Clearly, there are many pairs of leaves separated by $r$ from each other (cp.\ Figure \ref{fig-binarytree}(a)). Let $\{\ell_{n}(r)\}$ denote the set of all corresponding leaf-to-leaf distances. We now want to calculate the average leaf-to-leaf distance ${\cal L}_{n}(r)$ from the set $\{\ell_{n}(r)\}$. We first note that for a level $n$ tree the number of possible paths with separation $r$ is $2^{n}-r$.
In Figure \ref{fig-binarytree-maximalpaths}(b), we see that any complete level $n$ tree can be decomposed into two level $n-1$ sub-trees each of which contains $2^{n-1}$ leaves.
Let ${\cal S}_n(r)$ denote the sum of all possible leaf-to-leaf distances encoded in the set $\{\ell_{n}(r)\}$.
The structure of the decomposition in Figure \ref{fig-binarytree-maximalpaths}(b) suggests that we need to distinguish two classes of separations $r$. First, for $r<2^{n-1}$, paths are either completely contained within each of the two level $n-1$ trees or they bridge from the left level $n-1$ tree to the right level $n-1$ tree. Those which are completely contained sum to $2{\cal S}_{n-1}(r)$.
For those paths with separation $r$ that bridge across the two level $(n-1)$ trees, there are $r$ of such paths and each path has lengths $\ell_{n-1}=2n$.
Next, for $r\geq 2^{n-1}$, paths no longer fit into a level $n-1$ tree and always bridge from left to right. Again, each such path is $2n$ long and there are $L-r=2^{n}-r$ such paths.
Putting it all together, we find that
\begin{equation}
{\cal S}_{n}(r) = \left\{ \begin{array}{l l}
2{\cal S}_{n-1}(r) + 2nr, & \quad r < 2^{n-1}, \\
2n(2^{n} - r), & \quad r \geq 2^{n-1}.
\end{array} \right.
\label{eqn-rec-sumpathlength}
\end{equation}
for $n>1$ and with ${\cal S}_1(r)=1$.
%
%This recursive expression can be understood readily when looking at the binary structure of the tree. Clearly, the sum ${\cal S}_{n+1}(r)$ of all path lengths $\ell_{n+1}(r)$ at level $n+1$ will consist of the sum of path lengths for two level $n$ trees, plus the sum of all paths that connect the nodes across the two trees of level $n$. These paths have lengths $2n+1$ and for $r\leq 2^{n}$ there are $r$ such paths for the $r$ leaves while for $r>2^{n}$ there are $L-r = 2^{n+1}-r$ such paths.
%
Dividing by the total number of possible paths with separation $r$ then gives the desired average leaf-to-leaf distance
\begin{equation}
{\cal L}_{n}(r) \equiv \frac{{\cal S}_{n}(r)}{2^n-r}.
\label{eqn-rec-avgpathlength}
\end{equation}

%%%%%%%%%%%%%%%%%%%%%%%%%%%%%%%%%%%%%%%%%%%%%%%%%%%%%%%%%%%%%%%%%%%
\subsection{An explicit expression}
\label{sec-binary-averagepathlength-expression}
%%%%%%%%%%%%%%%%%%%%%%%%%%%%%%%%%%%%%%%%%%%%%%%%%%%%%%%%%%%%%%%%%%%

As long as $r < 2^{n-1}$, equation (\ref{eqn-rec-sumpathlength}) can be recursively expanded, i.e.\
\begin{subequations}
\begin{align}
\label{eqn-ana-Sexpand}
{\cal S}_{n}(r) & = 2{\cal S}_{n-1}(r) + 2nr\\
                & = 2 \left[ 2{\cal S}_{n-2}(r) + 2(n-1)r \right] + 2nr \\
                & = \ldots \nonumber
\end{align}
\end{subequations}
After $\nu$ such expansions, we arrive at
\begin{equation}
{\cal S}_{n}(r) = 2^{\nu} {\cal S}_{n-\nu}(r) + \sum_{k=0}^{\nu-1} 2^{k+1} (n-k) r.
\label{eqn-ana-Snsum}
\end{equation}
The expansion can continue while $r < 2^{n-\nu-1}$. It terminates when $n-\nu$ becomes so small such that the leaf separation $r$ is no longer contained within the level-$(n-\nu)$ tree. Hence the smallest permissible value of $n-\nu$ is given by \begin{equation}
n_c (r)= \lfloor \log_2 r \rfloor +1 ,
\label{eqn-ana-nc}
\end{equation}
where $\lfloor \cdot \rfloor$ denotes the floor function.
For clarity, we will suppress the $r$ dependence, i.e.\ we write $n_c\equiv n_c(r)$ in the following.
Continuing with the expansion of ${\cal S}_n(r)$ up to the $n_{c}$ term, we find
\begin{subequations}
\begin{align}
{\cal S}_{n}(r) & = 2^{n-n_c} {\cal S}_{n_c}(r) + \sum_{k=0}^{n-n_c-1} 2^{k+1}(n-k)\ r \label{eqn-ana-Sn-sum-1}\\
                & = 2^{n-n_c} {\cal S}_{n_c}(r) + [2^{n-n_c+1}(n_c+2)-2(n+2)]\ r \, .
\label{eqn-ana-Snrec}
\end{align}
\end{subequations}
Details for the summations occurring in Equation (\ref{eqn-ana-Snrec}) are given in Appendix \ref{sec-series}.
From Equation (\ref{eqn-rec-sumpathlength}), we have ${\cal S}_{n_c}(r) =2n_c(2^{n_c} - r)$, so equation (\ref{eqn-ana-Snrec}) becomes
\begin{equation}
{\cal S}_{n}(r)= 2^{n+1} ( n_c + 2^{1-n_c}r ) -2(n+2)r \, .
\label{eqn-ana-SnN}
\end{equation}
Hence the average leaf-to-leaf distances are given by
\begin{equation}
{\cal L}_n(r)= \frac{2}{2^n-r}\left[ 2^{n} ( n_c + 2^{1-n_c}r ) -(n+2)r \right] .
\label{eqn-ana-Lq}
\end{equation}
%%%%%%%%%%%%%%%%%%%%%%%%%%%%%%%%%%%%%%%%%%%%%%%%%%%%%%%%%%%%%%%%%%%
\begin{figure*}[tb]
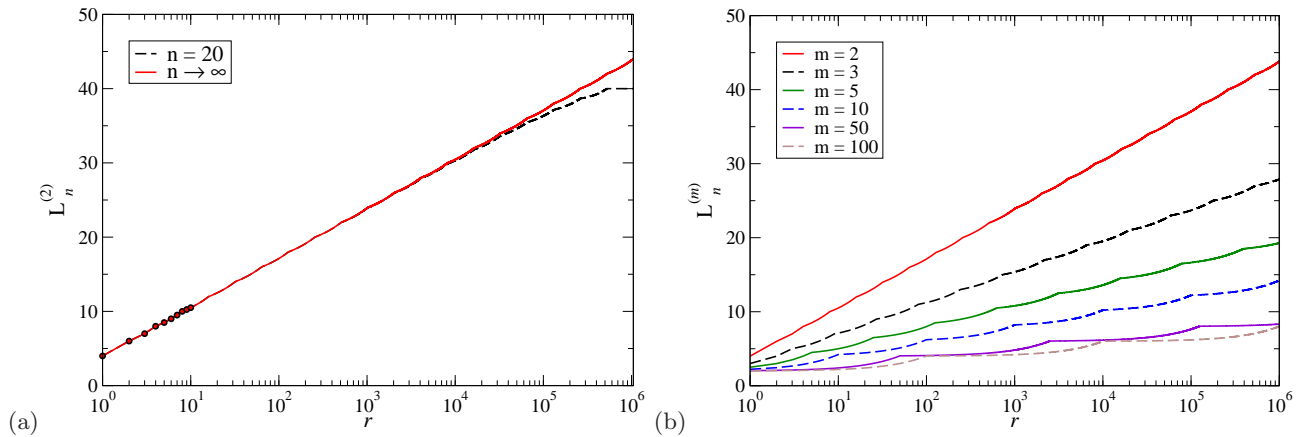

(a)\includegraphics[width=0.45\textwidth,clip]{n20tree.eps}
(b)\includegraphics[width=0.45\textwidth]{lim_m_1048575.eps}
\caption{(a) The average leaf-to-leaf distance ${\cal L}_n(r)$ versus leaf separation $r$ for a complete binary tree of $n= 20$ (dashed), i.e.\ length $L=2^{20}=1,048,576$, and also for $n\rightarrow\infty$ (solid). The first $10$ values are indicated by circles. (b) Average leaf-to-leaf distance ${\cal L}^{(m)}_{\infty}(r)$ for $m$-ary trees of various $m$. The curves for $m= 2, 5, 50$ are shown as solid lines, while those for $m=3, 10$ and $100$ have been indicated as dashed lines for clarity.}
\label{fig-bin-Lr}
\end{figure*}
%%%%%%%%%%%%%%%%%%%%%%%%%%%%%%%%%%%%%%%%%%%%%%%%%%%%%%%%%%%%%%%%%%%%
In the limit of $n \rightarrow\infty$ for fixed $r$, we have
\begin{equation}
\lim_{n\rightarrow \infty} {\cal L}_n(r)\equiv {\cal L}_{\infty}(r)= 2 \left( n_c + 2^{1-n_c}r \right).
\label{eqn-ana-Lq-inf}
\end{equation}
We emphasize that ${\cal L}_{\infty}(r) < \infty$ $\forall\ r < \infty$.

In Figure \ref{fig-bin-Lr}(a) we show finite and infinite leaf-to-leaf distances ${\cal L}_n(r)$. We see that whenever $r = 2^i$, $i \in \mathbb{N}$, we have a cusp in the ${\cal L}_n(r)$ curves. Between these points, the $\lfloor \cdot \rfloor$ function enhances deviations from the leading $\log_2 r$ behavior. This behavior is from the self-similar structure of the tree. Consider a sub-tree with $\nu$ levels, the largest separation that can occur in that sub-tree is $r = 2^{\nu}$, which has average distance $2\nu$. When $r$ becomes larger than the sub-tree size the leaf-to-leaf distance can no longer be $2\nu-1$ but always larger, so there is a cusp where this distance is removed from the possibilities. The constant average distance when $r \geq \frac{L}{2}$ is because there is only one possible leaf-to-leaf distance that connects the two primary sub-trees, which is clear from (\ref{eqn-rec-sumpathlength}).

%%%%%%%%%%%%%%%%%%%%%%%%%%%%%%%%%%%%%%%%%%%%%%%%%%%%%%%%%%%%%%%%%%%
\section{Generalization to complete m-ary trees}
\label{sec-general-averagepathlength}
%%%%%%%%%%%%%%%%%%%%%%%%%%%%%%%%%%%%%%%%%%%%%%%%%%%%%%%%%%%%%%%%%%%

%%%%%%%%%%%%%%%%%%%%%%%%%%%%%%%%%%%%%%%%%%%%%%%%%%%%%%%%%%%%%%%%%%%
\subsection{Average leaf-to-leaf distance in complete ternary trees}
\label{sec-ternary-averagepathlength}

Ternary trees are those where each node has \emph{three} children. Let us denote by ${\cal S}^{(3)}_n(r)$ and ${\cal L}^{(3)}_n(r)$ the sum and average, respectively, of all possible leaf-to-leaf distances $\{ \ell^{(3)}_n(r)\}$ for given $r$ in analogy to the binary case discussed before. Furthermore, $L=3^n$. Following the arguments which led to Equation (\ref{eqn-rec-sumpathlength}), we have
\begin{equation}
{\cal S}^{(3)}_{n}(r) = \left\{ \begin{array}{l l}
3{\cal S}^{(3)}_{n-1}(r) + 4nr, & \quad r < 3^{n-1} ,\\
2n(3^{n} - r), & \quad r \geq 3^{n-1}.
\end{array} \right.
\label{eqn-ternary-rec-sumpathlength}
\end{equation}
This recursive expression can again be understood readily when looking at the  structure of a ternary tree. Clearly, ${\cal S}^{(3)}_{n}(r)$ will now consist of the sum of leaf-to-leaf distances for three level $n$ trees, plus the sum of all paths that connect the nodes across the three trees of level $n$. The distances of these paths is solely determined by $n$ irrespective of the number of children and hence remains $2n$. As before, we need to distinguish between the case when $r$ fits within a level $n-1$ tree, i.e. $r< 3^{n-1}$, and when it connects different level $n-1$ trees, $r\geq 3^{n-1}$. For $r< 3^{n-1}$, there are now $2 r$ such paths, i.e., $r$ between the left and center level $n-1$ trees and $r$ the center and right level $n-1$ trees. For $r\geq 3^{n-1}$ there are $L-r = 3^{n}-r$ paths.
We again expand the recursion (\ref{eqn-ternary-rec-sumpathlength}) and find, with $n^{(3)}_{c} = \lfloor \log_{3}r \rfloor +1$ in analogy to (\ref{eqn-ana-nc}), that
\begin{equation}
{\cal S}^{(3)}_n(r)= 3^{n} \left[ 2n_{c}^{(3)} + 3^{1-n_{c}^{(3)}}r \right]  - (2n+3)r
\label{eqn-ternary-ana-Sq}
\end{equation}
and
\begin{align}
{\cal L}^{(3)}_n(r) &= \frac{S^{(3)}_n(r)}{3^n-r},\label{eqn-ternary-ana-Lq}\\
{\cal L}^{(3)}_{\infty}(r) &= 2n_{c}^{(3)} + 3^{1-n_{c}^{(3)}}r.\label{eqn-ternary-ana-Lq-inf}
\end{align}
%Here we have used $n^{(3)}_{c} = \lfloor \log_{3}r \rfloor +1$.

%%%%%%%%%%%%%%%%%%%%%%%%%%%%%%%%%%%%%%%%%%%%%%%%%%%%%%%%%%%%%%%%%%%
\subsection{Average leaf-to-leaf distance in complete $m$-ary trees}
\label{sec-mary-averagepathlength}
The methodology and discussion of the binary and ternary trees can be generalized to trees of $m>1$ children, known as \textit{$m$-ary} trees. The maximal leaf-to-leaf distance for any tree is independent of $m$ and determined entirely by the geometry of the tree. Each leaf node is at depth $n$, a maximal path has the root node as the lowest common ancestor, therefore the maximal path is $2n$.

A recursive function can be obtained using similar logic to before. For a given $n$, there are $m$ subgraphs with the structure of a tree with $n-1$ levels. When $r$ is less than the size of each subgraph ($r < m^{n-1}$), the sum of the paths is therefore the sum of $m$ copies of the subgraph along with the paths that connect neighboring pairs. When $r$ is larger than the size of the subgraph ($r \geq m^{n-1}$), the paths are all maximal. When all this is taken into account the recursive function is
\begin{equation}
{\cal S}^{(m)}_{n}(r) = \left\{ \begin{array}{l l}
m{\cal S}^{(m)}_{n-1}(r) + 2(m-1)nr, & \quad r < m^{n-1}, \\
2n(m^{n} - r), & \quad r \geq m^{n-1}.
\end{array} \right.
\label{eqn-mary-rec-sumpathlength}
\end{equation}
This can be solved in the same way as the binary case to obtain an expression for the sum of the paths for a given $m$, $n$ and $r$
\begin{equation}
\mathcal{S}^{(m)}_{n}(r) = 2 m^{n} \left[ n_{c}^{(m)} + \frac{m^{1-n_{c}^{(m)}}r}{(m-1)} \right] - 2r \left( n + \frac{m}{m-1} \right),
\label{eqn-mary-ana-Sq}
\end{equation}
The average leaf-to-leaf distance is then
\begin{equation}
{\cal L}^{(m)}_{n}(r) = \frac{{\cal S}^{(m)}_{n}(r)}{m^{n}-r}.
\label{eqn-mary-ana-Lq}
\end{equation}
and
\begin{equation}
%\lim_{n \to \infty} A_{n}(r) = m^{-\rho^{(m)}-1} \left[ (2\rho^{(m)}+1)(m^{\rho^{(m)}+1}-r) + \left[ 2\rho^{(m)}+3-\frac{2}{1-m} \right] r \right].
{\cal L}^{(m)}_{\infty}(r) = 2 \left[ n_{c}^{(m)} + \frac{m^{1-n_{c}^{(m)}}r}{(m-1)} \right] .
\label{eqn-mary-ana-Lq-inf}
\end{equation}
We note that in analogy with Equation (\ref{eqn-ana-nc}), we have used
\begin{equation}
n^{(m)}_{c}= \lfloor \log_m r \rfloor + 1
\end{equation}
in deriving these expressions.
Figure \ref{fig-bin-Lr}(b) shows the resulting leaf-to-leaf distances in the $n\rightarrow\infty$ limit for various values of $m$.
%
%We also have $\lim_{r \rightarrow\infty}{\cal L}^{(m)}_n(r)= 2n+\frac{2m}{m-1}$.

%%%%%%%%%%%%%%%%%%%%%%%%%%%%%%%%%%%%%%%%%%%%%%%%%%%%%%%%%%%%%%%%%%%
\section{Moments of the leaf-to-leaf distance distribution in complete m-ary trees}
%%%%%%%%%%%%%%%%%%%%%%%%%%%%%%%%%%%%%%%%%%%%%%%%%%%%%%%%%%%%%%%%%%%

%%%%%%%%%%%%%%%%%%%%%%%%%%%%%%%%%%%%%%%%%%%%%%%%%%%%%%%%%%%%%%%%%%%
\subsection{Variance of leaf-to-leaf distances in complete $m$-ary trees}
\label{sec-mary-variance}

In addition to the average leaf-to-leaf distance ${\cal L}^{(m)}_{n}(r)$, it is also of interest to ascertain its variance $\textrm{var}[{\cal L}^{(m)}_{n}](r)= \langle [{\cal L}^{(m)}_{n}(r)]^2 \rangle  - [{\cal L}^{(m)}_{n}(r)]^2$. Here $\langle \cdot \rangle$ denotes the average over all paths for given $r$ in an $m$-ary tree as before. In order to obtain the variance, we obviously need to obtain an expression for the sum of the squares of leaf-to-leaf distances. This can again be done recursively, i.e.\ with ${\cal Q}^{(m)}_n(r)$ denoting this sum of squared leaf-to-leaf distance for an $m$-ary tree of leaf separation $r$, we have similarly to Equation (\ref{eqn-mary-rec-sumpathlength})
\begin{equation}
{\cal Q}^{(m)}_{n}(r) = \left\{ \begin{array}{l l}
m{\cal Q}^{(m)}_{n-1}(r) + (m-1)4n^2r, & \quad r < m^{n-1}, \\
4n^2(m^{n} - r), & \quad r \geq m^{n-1} .
\end{array} \right.
\label{eqn-mary-rec-sumpathlengthsquare}
\end{equation}
Here, the difference to Equation (\ref{eqn-mary-rec-sumpathlength}) is that we have squared the distance terms $2n$. As before, expanding down to $n_c$ (here and in the following, we suppress the $(m)$ superscript of $n_{c}^{(m)}$ for clarity) gives a term containing ${\cal Q}^{(m)}_{n_c}(r)$,
%\begin{widetext}
%\begin{subequations}
%\begin{align}
%{\cal Q}^{(m)}_{n}(r) & = m^{n-n_c} {\cal Q}^{(m)}_{n_c}(r) + \sum_{k=0}^{n-n_c-1} 4(m-1)(n-k)^2 m^k r \\
%& = m^{n-n_c} {\cal Q}^{(m)}_{n_c}(r) + 4r(m-1) \sum_{k=0}^{n-n_{c}-1} \left[ n^{2} m^{k} - 2nk m^{k} + k^{2} m^{k} \right] \label{eqn-mary-Qn-sum-2}\\
%& = \frac{4}{(m-1)^{2}} \left\{ r m^{n-n_{c}+1} \left[m + 2n_{c}(m-1) + 1 \right] + m^{n}(m-1)^{2}n_{c}^{2} \right. \nonumber \\
%& \qquad \left. - r \left[ n^{2} + m^2(n+1)^{2} + m(1-2n(n+1)) \right] \right\} .
%\label{eqn-mary-ana-Qnrec}
%\end{align}
%\end{subequations}
%\end{widetext}
\begin{subequations}
\begin{align}
{\cal Q}^{(m)}_{n}(r) & = m^{n-n_c} {\cal Q}^{(m)}_{n_c}(r) + \nonumber \\ & \qquad \sum_{k=0}^{n-n_c-1} 4(m-1)(n-k)^2 m^k r \\
& = m^{n-n_c} {\cal Q}^{(m)}_{n_c}(r) + \nonumber \\ & \qquad 4r(m-1) \sum_{k=0}^{n-n_{c}-1} \left[ n^{2} m^{k} - 2nk m^{k} + k^{2} m^{k} \right] \label{eqn-mary-Qn-sum-2}\\
& = \frac{4}{(m-1)^{2}} \left\{ r m^{n-n_{c}+1} \left[m + 2n_{c}(m-1) + 1 \right] - \right. \nonumber \\
& \qquad \left. r \left[ n^{2} + m^2(n+1)^{2} + m(1-2n(n+1)) \right] + \right. \nonumber \\ & \left. \qquad \qquad m^{n}(m-1)^{2}n_{c}^{2} \right\} .
\nonumber \\
\label{eqn-mary-ana-Qnrec}
\end{align}
\end{subequations}
As before, details for the summations occurring in Equation (\ref{eqn-mary-Qn-sum-2}) are given in Appendix \ref{sec-series}.
We can therefore write for the variance
\begin{align}
\textrm{var}[{\cal L}^{(m)}_{n}](r)
&=\frac{{\cal Q}^{(m)}_n(r)}{m^n -r} -  \left[ {\cal L}^{(m)}_n(r) \right]^2 \nonumber \\
&=\frac{{\cal Q}^{(m)}_n(r)}{m^n -r} - \left[ \frac{{\cal S}^{(m)}_n(r)}{m^n -r} \right]^2 .
\end{align}
Using Equations (\ref{eqn-mary-ana-Qnrec}), (\ref{eqn-mary-ana-Lq}) and (\ref{eqn-mary-ana-Sq}), we then have explicitly
\begin{widetext}
\begin{align}
\textrm{var}[{\cal L}^{(m)}_{n}](r) & = \frac{4r}{m^{2n_{c}-2}(m^{n}-r)^{2}(m-1)^{2}} \Bigl( m^{2n} \Bigl[ m^{n_{c}-1}(m+1)-r \Bigr] + m^{2n_{c}-1}r  -
 m^{n} \Bigl\{ m^{n_{c}-1}(2n-2n_{c}+1)(m-1)r - \nonumber \\
& \qquad \quad m^{2n_{c}-2}(n_{c}-n)^{2} + m^{2n_{c}}(n-n_{c}+1)^{2} -
 m^{2n_{c}-1}\Bigl[2n^{2}-n(4n_{c}-2)+2n_{c}(n_{c}-1)-1\Bigr] \Bigr\} \Bigr),
\end{align}
\end{widetext}
and also
\begin{equation}
\textrm{var}[{\cal L}^{(m)}_{\infty}](r)
= \frac{4r \left[ m^{n_{c}-1} (m+1) -r \right]}{m^{2 n_{c}-2} (m-1)^2}.
\end{equation}
%%%%%%%%%%%%%%%%%%%%%%%%%%%%%%%%%%%%%%%%%%%%%%%%%%%%%%%%%%%%%%%%%%%
\begin{figure*}[tb]
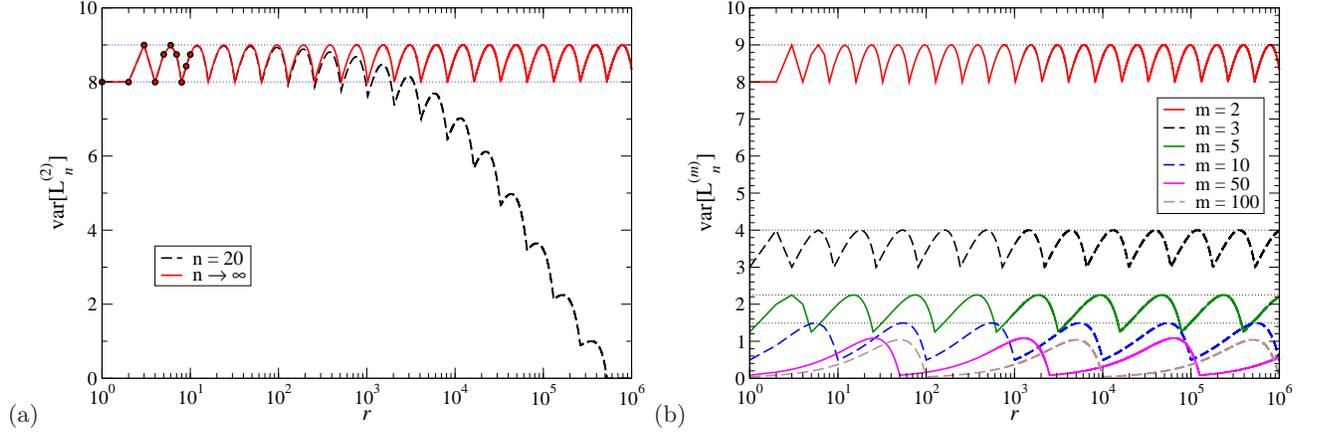

\centering
(a)\includegraphics[width=0.45\textwidth]{m2_n20_var_fin_in.eps}
(b)\includegraphics[width=0.45\textwidth]{lim_m_1048575_variance.eps}
\caption{Variance $\textrm{var}[{\cal L}^{(2)}_{n}](r)$ of the leaf-to-leaf distance for (a) binary trees. The two lines compare a finite tree ($n= 20$, dashed line) to an infinite tree (solid line). The circles indicate the first $10$ $\textrm{var}[{\cal L}^{(2)}_{n}]$ values similar to Figure \ref{fig-bin-Lr}(a). The two dotted horizontal lines correspond to $\textrm{var}[{\cal L}^{(m)}_{n}]= 8$ and $9$. (b)  $\textrm{var}[{\cal L}^{(m)}_{n}](r)$ for various $m$-ary trees indicates by  lines as in Figure \ref{fig-bin-Lr}(b). The $4$ dotted horizontal lines correspond to $\textrm{var}[{\cal L}^{(m)}_{\infty}]= 9, 4, 2.25, 1.49$.}
\label{fig-mary-moments-lim_m}
\end{figure*}
%%%%%%%%%%%%%%%%%%%%%%%%%%%%%%%%%%%%%%%%%%%%%%%%%%%%%%%%%%%%%%%%%%%
When $r= m^i$, $i \in \mathbb{N}^{0}$, then $\textrm{var}[{\cal L}^{(m)}_{\infty}]$ has a local minima and we find that $\textrm{var}[{\cal L}^{(m)}_{\infty}](m^i)= \frac{4 m}{(m-1)^2}$. Similarly, it can be shown that the local maxima are at $r=\frac{1}{2}m^{i}(m+1)$, then $\textrm{var}[{\cal L}^{(m)}_{\infty}]= \frac{4 m}{(m-1)^2} + 1$. These values are indicated in Figure \ref{fig-mary-moments-lim_m} for selected $m$.
%a local maxima in $\textrm{var}[{\cal L}^{(m)}_{\infty}]$ corresponds to those $r$ values where $\lfloor \log_{m} r \rfloor$ is furthest from an integer //dont think this is true

%%%%%%%%%%%%%%%%%%%%%%%%%%%%%%%%%%%%%%%%%%%%%%%%%%%%%%%%%%%%%%%%%%%
\subsection{General moments of leaf-to-leaf distances in complete $m$-ary trees}
\label{sec-mary-moments}

The derivation in section \ref{sec-mary-variance} suggests that any ${q}$-th raw moment of leaf-to-leaf distances can be calculated similarly as in Equation (\ref{eqn-mary-rec-sumpathlengthsquare}). Indeed, let us define ${\cal M}^{(m)}_{q,n}(r)$ as the ${q}$-th moment of an $m$-ary tree of level $n$ with leaf separation $r$. Then ${\cal M}^{(m)}_{1,n}(r)={\cal L}^{(m)}_{n}(r)$, ${\cal M}^{(m)}_{2,n}(r)={\cal Q}^{(m)}_{n}(r)$ and
\begin{equation}
\textrm{var}[{\cal L}^{(m)}_{n}](r)= \frac{{\cal M}^{(m)}_{2,n}(r)}{m^n -r} - \left[\frac{{\cal M}^{(m)}_{1,n}(r)}{(m^n -r)}\right]^2.
\end{equation}
Following Equation (\ref{eqn-mary-rec-sumpathlengthsquare}), we find
\begin{equation}
{\cal M}^{(m)}_{q,n}(r) = \left\{ \begin{array}{l l}
m{\cal M}^{(m)}_{q,n-1}(r) + 2^{q} n^{q} (m-1) r, & \quad r < m^{n-1}, \\
2^{q} n^{q} (m^{n} - r), & \quad r \geq m^{n-1} .
\end{array} \right.
\label{eqn-mary-rec-sumpathlengthmoment-recursion-a}
\end{equation}
By expanding, this gives
\begin{align}
{\cal M}^{(m)}_{q,n}(r) &=
 m^{n-n_c} {\cal M}^{(m)}_{q,n_c}(r) + \nonumber \\
 & \qquad \sum_{k=0}^{n-n_c-1} 2^{q} m^k (m-1) (n-k)^{q} r.
 \label{eqn-mary-rec-sumpathlengthmoment-nc}
\end{align}
As before, $n_c$ corresponds to the first $n$ value where, for given $r$, we have to use the second part of the expansion as in Equation (\ref{eqn-mary-rec-sumpathlengthmoment-recursion-a}). Hence we can substitute the second part of (\ref{eqn-mary-rec-sumpathlengthmoment-nc}) for ${\cal M}^{(m)}_{q,n_c-1}(r)$ giving
\begin{align}
{\cal M}^{(m)}_{q,n}(r) &=  m^{n-n_c} 2^{q} n_{c}^{{q}} (m^{n_c}-r) + \nonumber \\ & \qquad \sum_{k=0}^{n-n_c-1} 2^{q} m^k (m-1) (n-k)^{q} r.
\label{eqn-mary-rec-final}
\end{align}
In order to derive an explicit expression for this similar to section \ref{sec-binary-averagepathlength-expression}, we need again to study the final sum of Equation (\ref{eqn-mary-rec-final}). We write
\begin{widetext}
\begin{subequations}
\begin{align}
\sum_{k=0}^{n-n_{c}-1}2^q m^k (m-1) (n-k)^{q} r &= r(m-1)(-2)^{q} \left[ \sum_{k=0}^{\infty}m^{k} (k-n)^{{q}} - \sum_{k=n-n_{c}}^{\infty} m^{k} (k-n)^{{q}} \right] \\
\quad &= r(m-1)(-2)^{q} \left[ \sum_{k=0}^{\infty}m^{k} (k-n)^{{q}} - m^{n-n_{c}} \sum_{k=0}^{\infty} m^{k} (k - n_{c})^{{q}} \right]  \\
\quad &= r(m-1)(-2)^{q} \left[ \Phi \left( m,-{q},-n \right) -m^{n-n_{c}}\Phi \left( m,-{q},-n_{c} \right) \right] ,
\end{align}
\end{subequations}
\end{widetext}
where in the last step we have introduced the \emph{Hurwitz-Lerch Zeta function} $\Phi$ \cite{KanKY00,Sri13} (also referred to as the \emph{Lerch transcendent} \cite{GuiS08} or the \emph{Hurwitz-Lerch Transcendent} \cite{Mathematica9}). It is defined as the sum
\begin{equation}
\Phi(z,s,u) = \sum_{k=0}^{\infty} \frac{z^{k}}{(k+u)^{s}}, \quad z\in\mathbb{C}.
\end{equation}
The properties of $\Phi(z,s,u)$ are \cite{GuiS08}
\begin{subequations}
\begin{align}
\Phi(z,s,u+1) & = \frac{1}{z} \left[ \Phi(z,s,u)-\frac{1}{u^{s}} \right] \label{eqn:Phi_up1}, \\
\Phi(z,s-1,u) & = \left(u + z\frac{\partial}{\partial z} \right) \Phi(z,s,u) \label{eqn:Phi_sm1}, \\
\Phi(z,s+1,u) & = - \frac{1}{s} \frac{\partial \Phi}{\partial u} (z,s,u). \label{eqn:Phi_sp1}
\end{align}
\end{subequations}
Hence we can write
\begin{align}
{\cal M}^{(m)}_{q,n}(r) &=  m^{n-n_c} 2^q n_{c}^{{q}} (m^{n_c}-r) + \nonumber \\
& \qquad r(m-1)(-2)^{q} \left[ \Phi \left( m,-{q},-n \right) - \right. \nonumber \\
& \qquad \quad \left. m^{n-n_{c}}\Phi \left( m,-{q},-n_{c} \right) \right].
\label{eqn-mary-kmom}
\end{align}
Averages of ${\cal M}^{(m)}_{,n}(r) $ can be defined as previously via
\begin{equation}
{\cal A}^{(m)}_{q,n}(r) = \frac{{\cal M}^{(m)}_{{q},n}(r) }{m^n-r}
\end{equation}
such that ${\cal L}^{(m)}_{n}(r) = {\cal A}^{(m)}_{1,n}(r)$ and $\textrm{var}[{\cal L}^{(m)}_{n}](r)= {\cal A}^{(m)}_{2,n}(r) - \left[ {\cal A}^{(m)}_{1,n}(r) \right]^2$.

The properties (\ref{eqn:Phi_up1}) -- (\ref{eqn:Phi_sp1}) can be used to show that, for a given $m$ and ${q}$, $\Phi \left( m,-{q},-n \right)$ can be expressed as a polynomial of order $(-n)^{{q}}$. Therefore in the $n \to \infty$ limit, we find
\begin{align}
\lim_{n\rightarrow \infty} {\cal A}^{(m)}_{q,n}(r) &\equiv {\cal A}^{(m)}_{{q},\infty}(r) \nonumber \\
&= m^{-n_{c}} \left[ 2^q n_{c}^{{q}} (m^{n_c}-r) - \right. \nonumber \\
& \left. \qquad r(m-1)(-2)^{q} \Phi \left( m,-{q}, - n_{c} \right) \right]. \label{eqn-mary-kmom-limit}
\end{align}

%%%%%%%%%%%%%%%%%%%%%%%%%%%%%%%%%%%%%%%%%%%%%%%%%%%%%%%%%%%%%%%%%%%
\section{Complete $m$-ary trees with periodicity}
%%%%%%%%%%%%%%%%%%%%%%%%%%%%%%%%%%%%%%%%%%%%%%%%%%%%%%%%%%%%%%%%%%%

\begin{figure}[tb]%[tb]
\centering
\includegraphics[width=0.8\columnwidth,clip]{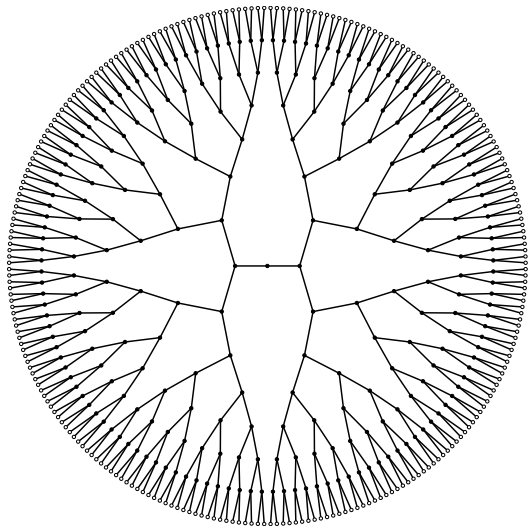}
\caption{A periodic, complete, binary tree with $n=8$ levels. Circles and lines as in Figure \ref{fig-binarytree}(a).}
\label{fig-periodicbinarytree}
\end{figure}

Up to now we have always dealt with trees in which the maximum separation $r$ was set by the number of leaves, i.e. $r \leq m^n$. This is know as a hard wall or \emph{open} boundary in terms of physical systems. A \emph{periodic} boundary can be realized by having the leaves of the tree form a circle as depicted in Figure \ref{fig-periodicbinarytree} for a binary tree.
%%%%%%%%%%%%%%%%%%%%%%%%%%%%%%%%%%%%%%%%%%%%%%%%%%%%%%%%%%%%%%%%%%%
%%%%%%%%%%%%%%%%%%%%%%%%%%%%%%%%%%%%%%%%%%%%%%%%%%%%%%%%%%%%%%%%%%%%
For such a binary tree, only separations $r \leq L/2$ are relevant since all cases with $r > L/2$ can be reduced to smaller $r=\mathrm{mod}(r,L/2)$ values by going around the periodic tree in the opposite direction. Therefore we can write
\begin{equation}
{\cal M}^{(m,\circ)}_{1,n}(r) = {\cal M}^{(m)}_{1,n}(r) + {\cal M}^{(m)}_{1,n}(m^{n}-r),
\label{eqn-pbc-opc}
\end{equation}
where $r < L/2$ and the superscript $\circ$ denotes the periodic case. Note that the case where $r = L/2$ the clockwise and anti-clockwise paths are the same so only need to be counted once. In the simple binary tree case we can expand this via (\ref{eqn-ana-SnN}) as in section \ref{sec-binary-averagepathlength-expression} and find
\begin{align}
{\cal M}^{(2,\circ)}_{1,n}(r) & \equiv{\cal S}^{(2,\circ)}_{n}(r) \nonumber \\
&= 2^{n+1}\left[ n_{c} + \tilde{n}_{c} -n -2 + \right. \nonumber \\
& \qquad \left. 2^{1-n_{c}}r + 2^{1-\tilde{n}_{c}}(2^n-r) \right],
\label{eqn-pbc-ana-Sq}
\end{align}
with $n_{c}$ as in Equation (\ref{eqn-ana-nc}) and $\tilde{n}_{c}= \lfloor \log_2 (2^n -r) \rfloor + 1$.
For every $r$, we have $2^n$ possible starting leaf positions on a periodic binary tree and hence the average leaf-to-leaf distance can be written as
\begin{align}
{\cal A}^{(2,\circ)}_{1,n}(r) &\equiv {\cal L}^{(2,\circ)}_n(r) = \frac{ {\cal S}^{(2,\circ)}_{n}(r) }{2^n} \nonumber \\
& = 2 \left[ n_{c} + \tilde{n}_{c} -n -2 + 2^{1-n_{c}}r + 2^{1-\tilde{n}_{c}}(2^n-r) \right] .
\label{eqn-pbc-ana-Lq}
\end{align}
This expression is the periodic analogue to Equation (\ref{eqn-ana-Lq}).
Generalizing to $m$-ary trees, with $\tilde{n}_{c} = \lfloor \log_m (m^n -r) \rfloor + 1$, we find
\begin{align}
{\cal M}^{(m,\circ)}_{1,n}(r) & = {\cal M}^{(m)}_{1,n}(r) + {\cal M}^{(m)}_{1,n}(m^{n}-r) \\
& = 2m^{n} \left[ n_{c}+\tilde{n}_{c} -n + \right. \nonumber \\
& \qquad \left. \frac{1}{m-1} \left( m^{1-n_{c}}r + m^{1-\tilde{n}_{c}}(m^{n}-r) -m \right) \right].
\end{align}
The average leaf-to-leaf distance for $m$-ary periodic trees is then given as
\begin{align}
{\cal A}^{(m,\circ)}_{1,n}(r)
&= \frac{{\cal M}^{(m,\circ)}_{1,n}(r) }{m^{n}} \nonumber \\
&= 2\left[ n_{c}+\tilde{n}_{c} -n + \right. \nonumber \\
&\qquad \left. \frac{1}{m-1} \left( m^{1-n_{c}}r + m^{1-\tilde{n}_{c}}(m^{n}-r) -m \right) \right] .
\label{eqn-pbc-av}
\end{align}

To again study the case of $n \to \infty$, it is necessary to observe how $\tilde{n}_{c}$ behaves for large $n$ and fixed $m$, $r$. When $n\gg r$, we have $r<m^{n-1}$ and hence $\lim_{n\rightarrow\infty}\lfloor \log_m (m^n -r) \rfloor = n-1$. This enables us to simply take the limits of Equation (\ref{eqn-pbc-av}) to give
\begin{equation}
\lim_{n\rightarrow \infty} {\cal A}^{(m,\circ)}_{1,n}(r) \equiv {\cal A}^{(m,\circ)}_{1,\infty}(r) = 2 \left[ n_{c} + \frac{m^{1-n_{c}}r}{(m-1)} \right],
\label{eqn-pbc-av-lim}
\end{equation}
which is the same as the open boundary case (\ref{eqn-mary-ana-Lq-inf}). This is to be expected as a small region of a large circle can be approximated by a straight line.

Last, the $q$-moments can be expressed similarly to Equation (\ref{eqn-mary-kmom}) via the Lerch transcendent as
\begin{align}
{\cal M}^{(m,\circ)}_{q,n}(r) & = {\cal M}^{(m)}_{q,n}(r) + {\cal M}^{(m)}_{{q},n}(m^{n}-r), \\
& = m^{n-n_c} 2^q n_{c}^{{q}} (m^{n_c}-r) + \nonumber \\
& \qquad m^{n-\tilde{n}_{c}} 2^{q} \tilde{n}_{c}^{{q}} (m^{\tilde{n}_{c}}-m^{n}+r) + \nonumber\\
& \qquad (m-1)(-2)^{q} \Big[ m^{n} \Phi(m,-q,-n) - \nonumber \\
& \qquad rm^{n-n_{c}} \Phi(m,-q,-n_{c}) - \nonumber\\
& \qquad (m^{n}-r)m^{n-\tilde{n}_{c}}\Phi(m,-q,-\tilde{n}_{c}) \Big],
\end{align}
The average $q$-moments in full are therefore
\begin{equation}
{\cal A}^{(m,\circ)}_{q,n}(r)  = \frac{{\cal M}^{(m,\circ)}_{{q},n}(r) }{m^{n}}
\end{equation}
for a complete, periodic, $m$-ary tree. To take the limit $n \to \infty$ notice that $\tilde{n}_{c} = n$ when $r<m^{n-1}$ for large $n$. Just like with Equation (\ref{eqn-pbc-av-lim}), this results in ${\cal A}^{(m,\circ)}_{q,\infty}(r) = {\cal A}^{(m)}_{{q},\infty}(r)$.

%%%%%%%%%%%%%%%%%%%%%%%%%%%%%%%%%%%%%%%%%%%%%%%%%%%%%%%%%%%%%%%%%%%
\section{Asymptotic scaling of the correlation for a homogeneous tree tensor network}
\label{sec-TTN}
%%%%%%%%%%%%%%%%%%%%%%%%%%%%%%%%%%%%%%%%%%%%%%%%%%%%%%%%%%%%%%%%%%%
% To illustrate where the properties calculated for complete trees may arise in tensor network simulations we aim to construct a tensor network wavefunction that has the same structure as the tree graphs under consideration throughout the paper.
% A tensor network that has the structure of a tree graph is known as a \emph{tree tensor network} (TTN) and is often used to model critical one dimensional quantum lattice systems due to the fact that they can be efficiently updated \cite{ShiDV06,TagEV09}.
%Tree tensor networks (TTNs) are often used to model critical one-dimensional many-body quantum lattice systems \cite{ShiDV06,TagEV09}. 
%The tensor network has the structure of a tree graph and can be efficiently updated.
Tree tensor networks (TTNs) are tensor networks that have the structure of a tree graph and are often used to model critical one-dimensional many-body quantum lattice systems as they can be efficiently updated \cite{ShiDV06,TagEV09}.
In principle it is possible to start from a tensor network wavefunction and derive a \emph{parent Hamiltonian} for which the wavefunction is a ground state \cite{VerWPC06,WolOVC06}.
In the case of homogeneous TTNs the procedure to create such a parent Hamiltonian seems likely to be highly non-trivial and not unique.
Here we build such a TTN from the \emph{binary} tree structure shown in Fig.\ \ref{fig-binarytree}.
At each internal vertex we place an isometric tensor \cite{EveV09,GolR14} with initially random entries and so-called bond dimension $\chi=4$.
Using as proxy a spin-1/2 Heisenberg model $H = \sum_{i=1}^{L-1} \vec{s}_{i} \cdot \vec{s}_{i+1}$, with $\vec{s}_{i}$ the spin-$1/2$ operator, we perform energy minimisation \cite{EveV09,TagEV09} at a bulk site. After each minimization, we replicate the bulk tensor to all other tensors such that every isometry is kept identical \cite{Sch11}. 
The process is then repeated until convergence (in energy). 
%We emphasize that we have chosen a low bond dimension $\chi=4$ such that the unphysical content of the tensors does not dominate the structure of the tree network \cite{Gol15}.
% 
% The ground state wavefunction for the Heisenberg model is known to be critical, but does not have the structure of a binary tree tensor network.
% Therefore we choose a low bond dimension so that the contents of the tensors do not mask the structure of the tree and we can more easily study the asymptotic correlation function.

\begin{figure}[tb]
\centering
\includegraphics[width=\columnwidth,clip]{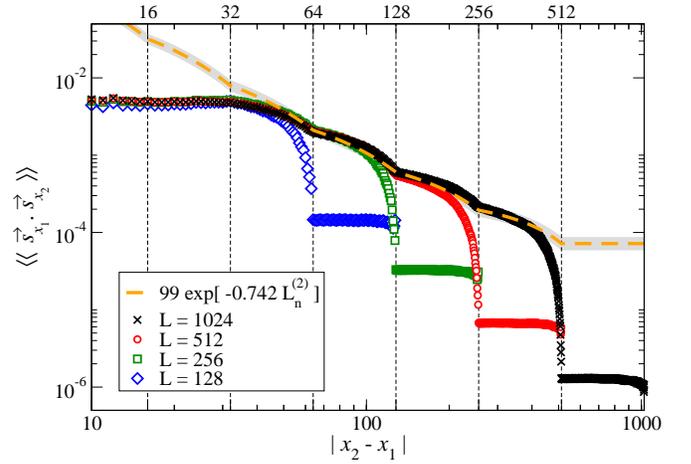}
\caption{ (Color Online)
Two point correlation function for TTNs with $\chi=4$ averaged over all pairs of sites separated by $|x_{2} - x_{1}|$ as discussed in text. 
The TTNs have $L = 128$ (blue diamonds), $256$ (green squares), $512$ (red circles), $1024$ (black crosses) corresponding to $n = 7$, $8$, $9$, $10$ levels respectively. 
The vertical dashed lines highlight $|x_{2} - x_{1}| = 16$, $32$, $64$, $128$, $256$, $512$.
%The orange dashed line corresponds to a power law fit of the asymptotic form of the correlation for $L = 1024$.
%The power is $-1.64 \pm 0.01$ and numerical prefactor is $2.06 \pm 0.09$.
The orange dashed line corresponds to a fit of $A \text{ exp}[-\alpha \mathcal{L}_{n}^{(2)}(r)]$ with $A = 99 \pm 9$ and $\alpha = 0.742 \pm 0.006$.
The grey shaded region is the standard error on the fit.
}
\label{fig-TTNcorrcomplete}
\end{figure}

A two-point correlation function $\langle \vec{s}_{x_{1}} \cdot \vec{s}_{x_{2}} \rangle$ is calculated \cite{TagEV09,GolR14} for all pairs of sites and averaged for all points separated by $|x_{2} - x_{1}|$.
The results are given in Fig.\ \ref{fig-TTNcorrcomplete}.
%%%%%%%%%%%%%%%%%%%%%%%%%%%%%%%%%%%%%%%%%%%%%%%%%%%%%%%%%%%%%%%%%%
%%%%%%%%%%%%%%%%%%%%%%%%%%%%%%%%%%%%%%%%%%%%%%%%%%%%%%%%%%%%%%%%%%%
For a homogeneous tensor network, a two-point correlation should scale as \cite{EveV11}
\begin{equation}
C(x_{1},x_{2}) \sim \text{exp}[-\alpha D_{TN}(x_{1},x_{2})],
\end{equation}
where $\alpha$ is a constant and $D_{TN}(x_{1},x_{2})$ is the number of tensor connecting sites $x_{1}$ and $x_{2}$.
Hence we expect the asymptotic correlation function to scale as $ \sim \text{exp} \left[ -\alpha \mathcal{L}_{n}^{(2)}(r) \right]$.
Figure \ref{fig-TTNcorrcomplete} shows that, away from small separations (e.g.\ $|x_{2} - x_{1}| > 32$ for $L=1024$), the content of the tensors no longer dominates the structural contribution and $\langle \vec{s}_{x_{1}} \cdot \vec{s}_{x_{2}} \rangle$ exhibits many of the properties we find in Fig.\ \ref{fig-bin-Lr}(a). The overall form of the long range correlations is a power law.
There are also the characteristic fluctuations from the self-similar structure of the tree with cusps at $|x_{2} - x_{1}| = 2^{i}$ for integer $i\geq 5$ (corresponding to $|x_{2} - x_{1}| > 32$).
When reaching the finite-size dominated regime $|x_{2}-x_{1}| \geq \frac{L}{2}$, we find an approximate constant average correlation. This is smaller than expected from Eq.\ (\ref{eqn-ana-Lq-inf}) because the top tensor of the TTN only has $\chi = 1$ and contributes less to the correlation function than the other tensors.
We emphasize that we have chosen a low bond dimension $\chi=4$ so that we can study the asymptotic form of the correlation functions for smaller system sizes.

The form of the correlations expressed in Fig.\ \ref{fig-TTNcorrcomplete} corresponds to those of a suitable parent Hamiltonian, i.e.\ one that has a ground state implied by this holographic tree structure. 
In addition, the results may also be useful for those building TTNs as a variational method for the study of critical systems.
The appearance of this form of the correlation for models that do not have a natural tree structure in the wavefunction, such as the Heisenberg model, is an indicator that the chosen $\chi$ is too small to capture the physics of the model. This is similar to the erroneous exponential decay of correlation functions found by DMRG for critical systems with power-law correlations in case of small $\chi$ \cite{Sch11}.
In all these situations the structure of the network dominates the value of the correlation rather than the information in the tensors. 

%%%%%%%%%%%%%%%%%%%%%%%%%%%%%%%%%%%%%%%%%%%%%%%%%%%%%%%%%%%%%%%%%%%
\section{Leaf-to-leaf distances for random binary trees}
\begin{figure}[tb]%[tb]
\centering
(a)\includegraphics[width=0.8\columnwidth,clip]{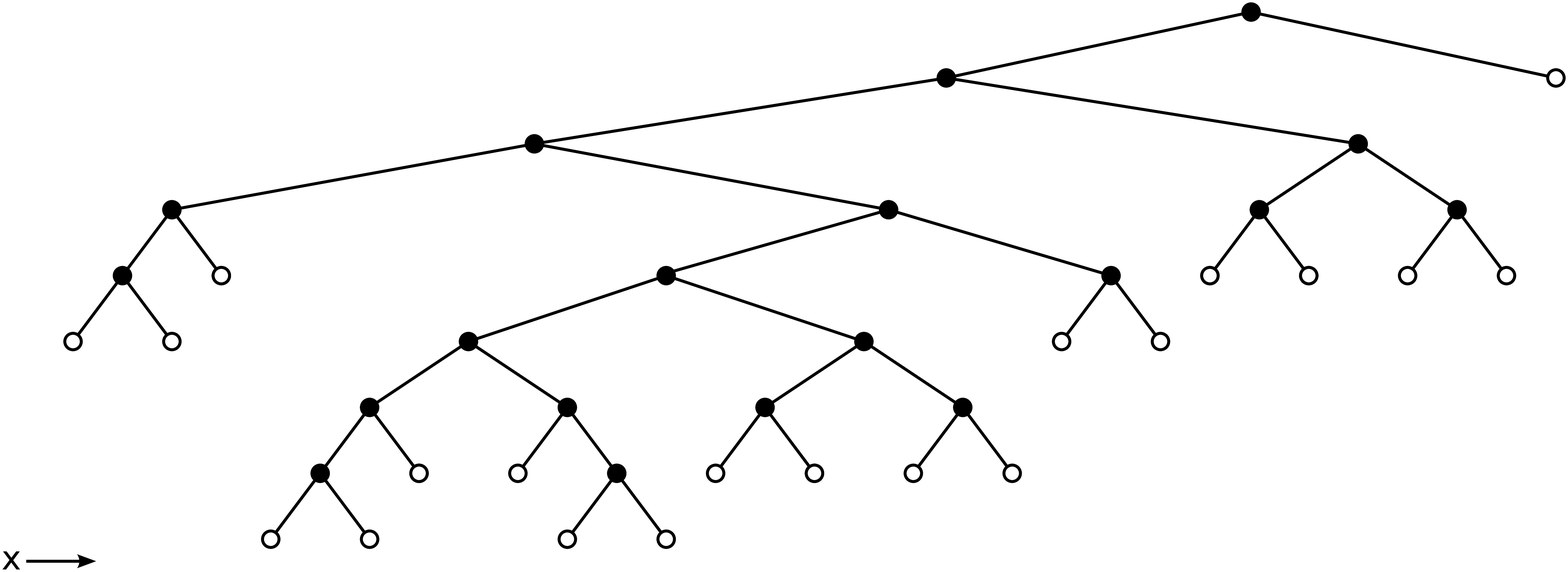}
(b)\includegraphics[width=0.8\columnwidth,clip]{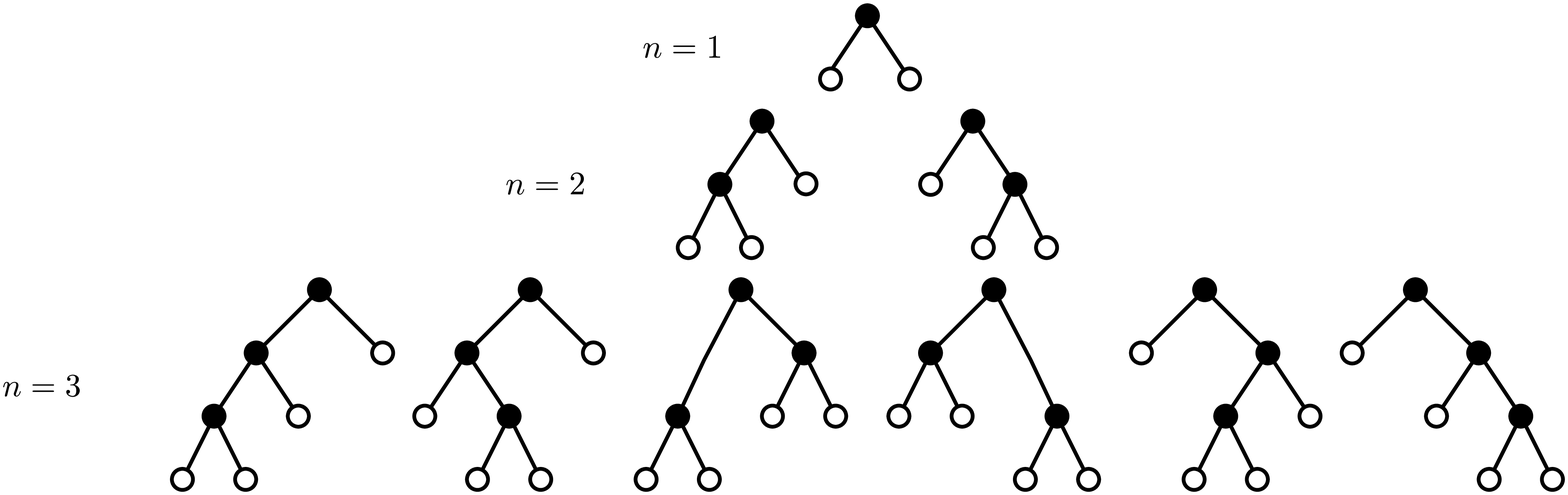}
\caption{(a) A random binary tree. (b) A complete set of random binary trees for $n=$ 1,2 and 3 ($L=2,3,4$). Circles and lines are as in Figure \ref{fig-binarytree}(a). }
\label{fig-randombinarytree}
\end{figure}

\begin{figure*}[tb]
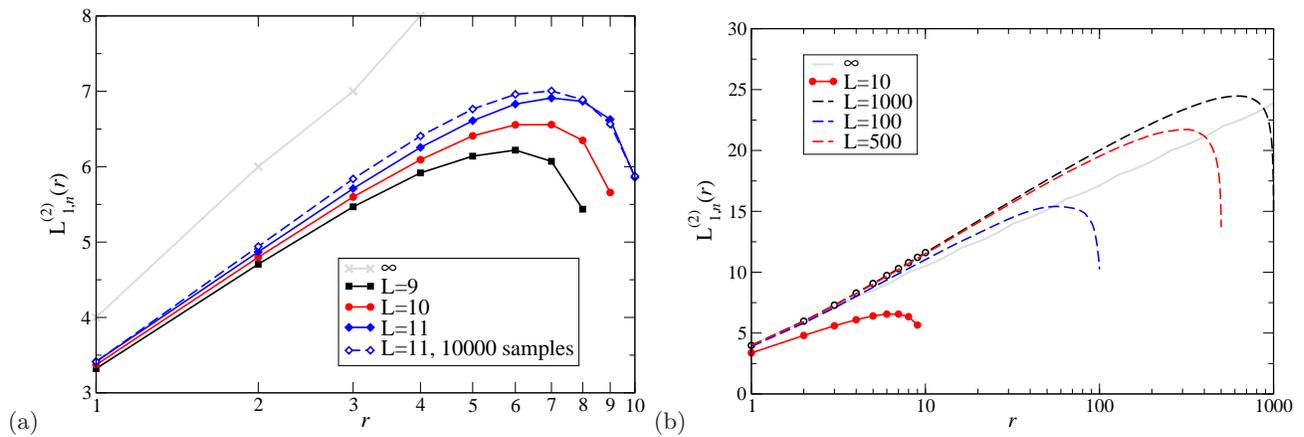

\centering
(a)\includegraphics[width=0.45\textwidth]{alltrees_9.eps}
(b)\includegraphics[width=0.45\textwidth]{randomtrees_1000_500.eps}
\caption{ (a) Average leaf-to-leaf distance through a random binary tree connecting two leaves of separation $r$ averaged over all possible trees for $L=9, 10$ and $11$ (solid symbols, lines are guide to the eye only). The open symbols (dashed line guide to the eye) refer to an average over $10000$ randomly chosen trees from the $10!$ possibilities for $L=11$. The grey crosses ($\times$) and line correspond to ${\cal L}^{(2)}_{\infty}(r)$ from Equation (\ref{eqn-ana-Lq-inf}). (b) Average leaf-to-leaf distance constructed from $500$ randomly chosen binary trees with $L=1000$ (dashed line). The open symbols ($\circ$) denotes the first $10$ data points. The closed symbols (red $\bullet$) and the solid line correspond to the $L=10$ data from (a). The grey line correspond to ${\cal L}^{(2)}_{\infty}(r)$ as in (a). Error bars have been omitted in (a) and (b) as they are within symbol size.}
\label{fig:random_1000_500}
\label{fig:alltrees_9}
\label{fig-binary-random}
\end{figure*}

%%%%%%%%%%%%%%%%%%%%%%%%%%%%%%%%%%%%%%%%%%%%%%%%%%%%%%%%%%%%%%%%%%%
In Figure \ref{fig-randombinarytree}(a) we show a binary tree where the leaves do not all appear at the same level $n$, but rather each node can become a leaf node according to an independent and identically distributed random process. Such trees are no longer complete, but nevertheless have many applications in the sciences \cite{SedF13,GolR14}.
%%%%%%%%%%%%%%%%%%%%%%%%%%%%%%%%%%%%%%%%%%%%%%%%%%%%%%%%%%%%%%%%%%%
%%%%%%%%%%%%%%%%%%%%%%%%%%%%%%%%%%%%%%%%%%%%%%%%%%%%%%%%%%%%%%%%%%%
%
Let us again compute the average leaf-to-leaf distance ${\cal L}^{(2,{\cal R})}_{n}(r)$ for a given $r$, when all possible pairs of leaves of separation $r$ and all possible trees of $L-1$ internal nodes are considered. Here ${\cal R}$ denotes the random character of trees under consideration.
For each $n=L-1$, there are $n!$ different such random trees as shown in Figure \ref{fig-randombinarytree}(b). We construct these trees numerically and measure ${\cal L}^{(2,{\cal R})}_{n}(r)$ as shown in Figure \ref{fig-binary-random}(a) \footnote{We emphasize that this definition of a random trees is different from the definition of so-called Catalan tree graphs \cite{SedF13}, as the number of unique graphs is given by the Catalan number $C_n$ and does not double count the degenerate graphs as shown in the center of the $n=3$ case of Figure \ref{fig-randombinarytree}(b).}.
For small $n$, we have computed all $(L-1)!$ trees (cp.\ Figure \ref{fig-binary-random}(a)) while for large $n$, we have averaged over a finite number $N\ll (L-1)!$ of randomly chosen binary trees among the $(L-1)!$ possible trees (cp.\ Figure \ref{fig-binary-random}(b)).
%%%%%%%%%%%%%%%%%%%%%%%%%%%%%%%%%%%%%%%%%%%%%%%%%%%%%%%%%%%%%%%%%%
%%%%%%%%%%%%%%%%%%%%%%%%%%%%%%%%%%%%%%%%%%%%%%%%%%%%%%%%%%%%%%%%%%%
We see in Figure \ref{fig-binary-random}(a) that, similar to the complete binary trees considered in the section \ref{sec-binary-averagepathlength}, the leaf-to-leaf distances increase with $r$ until they reach a maximal value. Unlike the complete tree in Figure \ref{fig-bin-Lr}(a), they start to decrease rapidly beyond this point. We also see that for such small trees, we are still far from the infinite complete tree result ${\cal L}^{(2)}_{\infty}(r)$ of Equation (\ref{eqn-ana-Lq-inf}). Finally, we also see that when we choose $10,000$ random binary trees from the $10!=3,628,800$ possible such trees at $L=11$ that the average leaf-to-leaf distances for each $r$ is still distinguishably different from an exact summation of all leaf-to-leaf distances. This suggests that rare tree structures are quite important.
In Figure \ref{fig-binary-random}(b) we nevertheless show estimates of ${\cal L}^{(2,{\cal R})}_{n}(r)$ for various $n$. As before, the shape of the curves for large $n$ is similar to those for small $n$. Clearly, however, the cusps in ${\cal L}^{(2)}_{n}(r)$ are no longer present in ${\cal L}^{(2,{\cal R})}_{n}(r)$. Also, the values of ${\cal L}^{(2,{\cal R})}_{n}(r)$ are larger than those for ${\cal L}^{(2)}_{n}(r)$ for small $r$.

%%%%%%%%%%%%%%%%%%%%%%%%%%%%%%%%%%%%%%%%%%%%%%%%%%%%%%%%%%%%%%%%%%%
\section{Conclusions}
%%%%%%%%%%%%%%%%%%%%%%%%%%%%%%%%%%%%%%%%%%%%%%%%%%%%%%%%%%%%%%%%%%%
We have calculated an analytic form for the average distance between two leaves with a given separation --- ordered according to the physical distance long a line --- in a complete binary tree graph. 
This result is then generalized to a complete tree where each vertex has any finite number of children. 
In addition to the mean leaf-to-leaf distance, it is found that the raw moments of the distribution of leaf-to-leaf distances have an analytic form that can be expressed in a concise way in terms of the Hurwitz-Lerch Zeta function. 
These findings are calculated for open trees, where the leaves form an open line, periodic trees, where the leaves form a circle, and infinite trees, which is the limit where the number of levels, $n$, goes to infinity. 
Each of these results has a concise form and characteristic features due to the self-similarity of the trees. 
We believe that these results provide a useful insight into the structure of the regular tree graphs that are relevant for the field of tensor networks \cite{SilGMR10,GerSRF14}. 
We also note that leaf-to-leaf distances computed here are qualitatively similar, but quantitatively different from those for random-spin chains \cite{GolR14}. 
This points to a subtle, yet physically relevant, difference in their Hilbert space properties.

%%%%%%%%%%%%%%%%%%%%%%%%%%%%%%%%%%%%%%%%%%%%%%%%%%%%%%%%%%%%%%%%%%%
\begin{acknowledgments}
We thank M.\ Bates, A.\ Czumaj and G.\ Rowlands for discussions. We would like to thank the EPSRC for financial support (EP/J003476/1) and
  provision of computing resources through the MidPlus Regional HPC Center (EP/K000128/1).
\end{acknowledgments}

%%%%%%%%%%%%%%%%%%%%%%%%%%%%%%%%%%%%%%%%%%%%%%%%%%%%%%%%%%%%%%%%%%%
\appendix

%%%%%%%%%%%%%%%%%%%%%%%%%%%%%%%%%%%%%%%%%%%%%%%%%%%%%%%%%%%%%%%%%%%
\section{Some useful series expressions}
\label{sec-series}
%%%%%%%%%%%%%%%%%%%%%%%%%%%%%%%%%%%%%%%%%%%%%%%%%%%%%%%%%%%%%%%%%%%

%%%%%%%%%%%%%%%%%%%%%%%%%%%%%%%%%%%%%%%%%%%%%%%%%%%%%%%%%%%%%%%%%%%
\subsection{Series used in section \ref{sec-binary-averagepathlength-expression}}

When the last sum in Equation (\ref{eqn-ana-Sn-sum-1}) is expanded, it is simply the sum of two geometric series. The first part can be simplified using
\begin{equation}
\sum_{k=1}^{l} x^{k} = \frac{x(1-x^{l})}{1-x},
\label{eqn:Pgeo1}
\end{equation}
the second part uses the arithmetico-geometric series
\begin{equation}
\sum_{k=1}^{l} kx^{k+1} = \frac{x(1-x^{l+1})}{(1-x)^2} - \frac{x+lx^{l+2}}{1-x}.
\label{eqn:Pgeo2}
\end{equation}

%%%%%%%%%%%%%%%%%%%%%%%%%%%%%%%%%%%%%%%%%%%%%%%%%%%%%%%%%%%%%%%%%%%
\subsection{Series used in section \ref{sec-mary-variance}}

The explicit expressions for the series terms occurring in Equation (\ref{eqn-mary-Qn-sum-2}) are given here. The first part is again a simple geometric series
%\begin{equation}
$\sum_{k=0}^{l} x^{k} = \frac{1-x^{l+1}}{1-x}$
%\label{eqn:Pgeo3}
%\end{equation}
similar to \ref{eqn:Pgeo1}. The second part is similar to (\ref{eqn:Pgeo2}),
%\begin{equation}
$\sum_{k=0}^{l} kx^{k} = \frac{x(1-x^{l})}{(1-x)^2} - \frac{lx^{l+1}}{1-x}$.
%\label{eqn:Pgeo4}
%\end{equation}
The final part is also an arithmetico-geometric series and has the form \cite{GraR94}
\begin{align}
\sum_{k=0}^{l-1} k^{2}x^{k} &= \frac{1}{(1-x)^{3}} \left[(-l^{2}+2l-1)x^{l+2}+ \right. \nonumber \\
& \qquad \left. (2l^{2}-2l-1)x^{l+1}-l^{2}x^{l}+x^{2}+x \right].
\end{align}

\end{document}